\documentclass[fleqn,usenatbib]{mnras}

\usepackage{newtxtext,newtxmath}

\usepackage[T1]{fontenc}
\usepackage{ae,aecompl}

\usepackage{amsfonts}
\usepackage{amsmath}
\usepackage{amssymb}
\usepackage{graphicx}
\usepackage{xcolor}
\usepackage{microtype}

\newcommand{\be}{\begin{equation}}
\newcommand{\ee}{\end{equation}}

\title[Bias:  Mass vs environment]
      {Dependence of halo bias on mass and environment} 

\author[J. Shi \& R. K. Sheth]
{Jingjing Shi$^{1,2}$\thanks{E-mail: jshi@sissa.it} \& Ravi K. Sheth$^{3}$\\  
 $^1$ SISSA - International School For Advanced Studies,
      Via Bonomea, 265 34136 Trieste, Italy\\
 $^2$ Department of Astronomy, University of Science and Technology of China, Hefei, 230026 Anhui, China\\
 $^3$ Centre for Particle Cosmology, University of Pennsylvania, 
      209 S. 33rd St., Philadelphia, PA 19104, USA\\
}

\date{Accepted XXX. Received YYY; in original form ZZZ}

\pubyear{2017}

\begin{document}
\label{firstpage}
\pagerange{\pageref{firstpage}--\pageref{lastpage}}
\maketitle

\begin{abstract}
The simplest analyses of halo bias assume that halo mass alone determines halo clustering.  However, if the large scale environment is fixed, then halo clustering is almost entirely determined by environment, and is almost completely independent of halo mass.  We show why.  Our analysis is useful for studies which use the environmental dependence of clustering to constrain cosmological and galaxy formation models. It also shows why many correlations between galaxy properties and environment are merely consequences of the underlying correlations between halos and their environments, and provides a framework for quantifying such inherited correlations.
\end{abstract}


\section{Introduction}\label{sec:intro}
The clustering of galaxies is often used to constrain models of the background cosmology and galaxy formation.  In many studies, the Halo Model \citep{Cooray2002} plays an important role.  In the simplest (and most widely used) version of the approach, the clustering of galaxies is determined by a combination of how galaxies populate halos, and the clustering of the halos which host galaxies, and for both ingredients, halo mass is assumed to be the only halo property which matters.  \citet{Abbas2007} describe one of the first tests of this assumption; they classified galaxies by the number of neighbours within $\sim 8h^{-1}$Mpc; measured the clustering signal as a function of environment; and showed that the environmental dependence of clustering was similar to that in a mock catalog in which the galaxy content of a halo was determined completely by halo mass and not environment.  

However, they also showed that they were able to model the strength of the clustering signal as a function of environment alone.  This finding has recently been confirmed by \citet{Pujol2017}. Provided that the environment is defined on a scale that is substantially larger than a typical halo, the clustering signal is a function of environment, and not of halo mass.  I.e., at fixed environment, the clustering is independent of halo mass, whereas at fixed mass, the clustering is a strong function of environment.  The main goal of the present paper is to provide a more careful derivation of the expression in \citet{Abbas2007}. A final section discusses how this particular clustering signal is related to what has come to be called Assembly Bias \citep{Sheth2004}, and makes the point that it is useful to distinguish between halo--environment correlations, and whether or not the way galaxies populate halos requires additional correlations.

\section{Bias of constrained regions}\label{Lag}
In what follows, it is important to distinguish clearly between the scale associated with halo formation, that on which the environment is defined, and the (typically much larger) scale on which the bias factor is measured.  We will use $R_h$, $R_e$ and $R_0$ to denote these scales.  

\subsection{Large scale environment as a constraint}
Suppose that we identify those positions in the initial (Gaussian) field which, when smoothed on scale $R_e$ have overdensity $\Delta_e$.  Let $S_e\equiv\langle\Delta_e^2\rangle$.  The probability of being centred on such a region is 
\be
 p(\Delta_e) = \frac{{\rm exp}(-\Delta_e^2/2S_e)}{\sqrt{2\pi S_e}}.
\ee
The conditional probability that the overdensity, when smoothed on some other scale $R_0$, is $\Delta_0$, given that it is $\Delta_e$ on scale $R_e$, is 
\be
 p(\Delta_0|\Delta_e) 
 = \frac{{\rm exp}^{-(\Delta_0-\mu_{0|e})^2/2S_{0|e}}}{\sqrt{2\pi S_{0|e}}},
\ee
where 
\be
 \mu_{0|e} = \langle\Delta_0|\Delta_e\rangle
    = \frac{\langle\Delta_0\Delta_e\rangle}{\langle\Delta_e^2\rangle}\,\Delta_e
\ee
and 
\be
 S_{0|e} \equiv S_0\,(1 - \langle\Delta_0\Delta_e\rangle^2/S_0S_e).
\ee
Now, $\langle\Delta_0\Delta_e\rangle$ is the correlation between $\Delta$ on the two scales, whereas $\langle\Delta_0|\Delta_e\rangle$ is the cross correlation between the two $\Delta$s subject to the constraint that $\Delta=\Delta_e$ on scale $R_e$.  Hence, it is natural to define  
\be
 \langle\Delta_0|\Delta_e\rangle \equiv b_e\,\langle\Delta_0\Delta_e\rangle
 \qquad{\rm where}\quad 
 b_e \equiv \Delta_e/S_e.
 \label{bKaiser}
\ee
The expression above shows that we should think of the constrained cross-correlation as biasing the unconstrained correlation; the bias is linearly proportional to the constraint.  At the risk of belaboring the point, 
\begin{align}
\langle\Delta_0\Delta_e\rangle 
 &= \int {\rm d}\Delta_e\,\int {\rm d}\Delta_0\,p(\Delta_e)\,p(\Delta_0|\Delta_e) \Delta_0\Delta_e \nonumber\\
 &= \int {\rm d}\Delta_e\,p(\Delta_e)\,\Delta_e \langle\Delta_0|\Delta_e\rangle \nonumber\\
 &= \int {\rm d}\Delta_e\,p(\Delta_e)\,\Delta_e \,b_e \langle\Delta_0\Delta_e\rangle \nonumber\\
 &= \langle\Delta_0\Delta_e\rangle \int {\rm d}\Delta_e\,p(\Delta_e)\,\frac{\Delta_e^2}{S_e} ,
\end{align}
where the final integral equals unity.  The second expression shows that the unconstrained correlation $\langle\Delta_0\Delta_e\rangle$ is a weighted sum over the constrained cross-correlations $\langle\Delta_0|\Delta_e\rangle$.  

Equation~(\ref{bKaiser}) for the bias is familiar in cosmology from \citet{Kaiser1984}. However, there it was introduced in the context of the bias associated with regions which exceed a high threshold in a Gaussian field, though it is often referred to as the `high peak' limit.  The expression above shows that this bias expression is actually associated with a much simpler constraint than either thresholds or peaks:  simply that the height {\em equals} a certain value. The correspondence with peaks is a consequence of the fact that if $\Delta_e\gg \sqrt{S_e}$, then the additional constraints which define a peak do not matter for the bias (because the highest positions in the field are almost certainly also local peaks).  While this reason was clear in early work, some more recent papers -- arguing that equation~(\ref{bKaiser}) is particular to peaks -- have got the logic backwards.  

\subsection{Small scale overdensity as an additional constraint}\label{sharpk}
The analysis of the previous section shows what one should expect if the constraints are more complicated.  E.g., if we add a constraint on a third scale $R_h$, then  
\begin{align}
 \langle\Delta_0|\Delta_e,\Delta_h\rangle 
 &= \langle\Delta_0|\Delta_e\rangle + \langle\Delta_0|\Delta_{h|e}\rangle
 \label{orth}
\end{align}
where 
\be
 \Delta_{h|e}\equiv \Delta_h - \langle\Delta_h|\Delta_e\rangle.
\ee

Notice that, as there are now two constraints, the bias is the sum of two terms; the form of the expression above suggests that we should think of the prefactors of the correlations with $\Delta_e$ and $\Delta_{h|e}$ as being two bias factors.  However, $\Delta_{h|e}$ involves both $\Delta_h$ and $\Delta_e$, whereas we are typically interested in keeping the effects of these two terms separate.  I.e., we seek the coefficients of the terms proportional to $\langle\Delta_0\Delta_e\rangle$ and $\langle\Delta_0\Delta_h\rangle$, respectively.  If we define $\nu_e\equiv \Delta_e/\sqrt{S_e}$, and similarly for $\nu_h$, then a little algebra shows that
\begin{equation}
\langle\Delta_0|\Delta_e,\Delta_h\rangle =
b_e\, \langle\Delta_0\Delta_e\rangle + b_h\,\langle\Delta_0\Delta_h\rangle
\end{equation}
where
\begin{align} 
 b_e &= \frac{\nu_e - \langle\nu_e\nu_h\rangle\, \nu_h}{\sqrt{S_e}(1 - \langle\nu_e\nu_h\rangle^2)} \\
 b_h &= \frac{\nu_h - \langle\nu_h\nu_e\rangle\, \nu_e}{\sqrt{S_h}(1 - \langle\nu_e\nu_h\rangle^2)} 
.
\end{align}
Note that $b_e\to \Delta_e/S_e$ only when $\langle\nu_e\nu_h\rangle\to 0$ and, in this limit, $b_h\to \Delta_h/S_h$ as well.  I.e., when $\langle\nu_e\nu_h\rangle\to 0$, both $b_e$ and $b_h$ have the form of equation~(\ref{bKaiser}) in their respective variables.  Typically, this will happen when $R_e\gg R_h$.  The approach to zero will be faster if $R_e$ and $R_h$ are centred on different positions.  Furthermore, if $R_0\gg R_e$ and $R_e\gg R_h$, then we expect $\langle\Delta_0\Delta_h\rangle\ll \langle\Delta_0\Delta_e\rangle$.  In this limit $\langle\Delta_0|\Delta_e,\Delta_h\rangle \to \langle\Delta_0|\Delta_e\rangle$:  the constraint on $\Delta_h$ is irrelevant.  

In the present context, equation~(\ref{orth}) is the more transparent expression because it shows that the constraint on $\Delta_h$ will be irrelevant if $\langle\Delta_0|\Delta_{h|e}\rangle=0$.  This happens if $\langle\Delta_0\Delta_h\rangle = \langle\Delta_0\Delta_e\rangle\langle\Delta_e\Delta_h\rangle/S_e$; i.e., if the $\Delta_0$-$\Delta_h$ correlation is entirely a consequence of the $\Delta_0$-$\Delta_e$ and $\Delta_e$-$\Delta_h$ correlations.  This holds true for the special case when the smoothing filter used to define $\Delta$ on the different scales is sharp in $k$-space.  Such a filter was used extensively in the past, as it leads to Markovian walks with uncorrelated steps, which renders many questions of interest analytically tractable \citep{Bond1991}.  For this filter, $\langle\Delta_r\Delta_R\rangle = \langle\Delta_R^2\rangle$ where $R\ge r$.  Hence, for this filter $\langle\Delta_0\Delta_h\rangle = S_0$ and $\langle\Delta_0\Delta_e\rangle\langle\Delta_e\Delta_h\rangle/S_e = S_0S_e/S_e = S_0$.  So, if $\Delta_e$ is fixed, then the constraint from $\Delta_h$ is {\em completely} irrelevant.  

More generally, the constraint on $\Delta_h$ will be irrelevant if $\langle\Delta_0|\Delta_{h|e}\rangle\ll \langle\Delta_0|\Delta_e\rangle$, i.e., if the amount of correlation between $\Delta_0$ and $\Delta_h$ which is not due the $\Delta_0$-$\Delta_e$ and $\Delta_e$-$\Delta_h$ correlations is smaller than the $\Delta_0$-$\Delta_e$ correlation.  When $R_0\gg R_e\gg R_h$, this is very likely to be the case.  Hence, except when $|\Delta_{h|e}|$ is very large the fact that $\Delta_h$ is constrained will not matter; the cross correlation $\langle\Delta_0|\Delta_e,\Delta_h\rangle$ will be dominated by the first term on the rhs of equation~(\ref{orth}).  Recalling that our choice of subscripts is not accidental, this discussion implies that when $R_e\gg R_h$ then the cross correlation signal of equation~(\ref{orth}) will be dominated by the correlation with the environment; the halo mass is almost always irrelevant.  Halo mass only matters if $|\Delta_{h|e}|$ is large: since $\Delta_h$ is typically of order unity, halo mass matters more if $\Delta_e$ is very negative (i.e. in underdense regions).  

\subsection{Small scale overdensity and its derivatives as additional constraints}

Equation~(\ref{orth}) serves mainly to illustrate how cross-correlations with the large scale environment generalize as one adds more constraints.  Following \citet{Musso2012}, we are most interested in the case in which the derivatives of $\Delta_h$ also matter.  In this case, 
\begin{align}
 \langle\Delta_0|\Delta_e,\Delta_h,\Delta'_h\rangle 
 &= \langle\Delta_0|\Delta_e\rangle + \langle\Delta_0|\Delta_{h|e}\rangle
                                   + \langle\Delta_0|\Delta_{h'|he}\rangle
\end{align}
and we are again faced with the problem of showing when the first term on the right hand side dominates.  

In this context, it is interesting to consider a slightly more general problem in which the derivative on scale $R_e$ is also specified.  Then we are interested in 
 $\langle\Delta_0|\Delta_e,\Delta'_e,\Delta_h,\Delta'_h\rangle$.  
\citet{Musso2014} describe a family of -- what they call Markov Velocity -- models in which correlations between scales are rather similar to those in $\Lambda$CDM models.  They show that, for Markov Velocity models,  
\be
 \langle\Delta_0|\Delta_e,\Delta'_e,\Delta_h,\Delta'_h\rangle 
 = \langle\Delta_0|\Delta_e,\Delta'_e\rangle ;
\ee
i.e., if both $\Delta_e$ and $\Delta'_e$ are specified, then the smaller scale $R_h$ is irrelevant (see their equation~71).  For Markov Velocity models this is an exact, not an approximate, statement.  As a result, $\langle\Delta_0|\Delta_e,\Delta_h,\Delta'_h\rangle$ only depends weakly on $R_h$, or depends on $R_h$ only for a rather restricted range of scales.
The similarity of these models to $\Lambda$CDM strongly suggests that $\langle\Delta_0|\Delta_e,\Delta_h,\Delta'_h\rangle$ in $\Lambda$CDM models will also only depend weakly on $R_h$.  I.e., if the environment on scale $R_e\ge R_h$ is fixed, then the large scale bias is approximately independent of halo mass.  

\subsection{General formulation}
The lesson from the previous explicit models is clear.  If the vector ${\bmath h}$ includes all the variables which are important for halo formation, then one should express halos as constraints on these variables in the underlying Gaussian field:
\be
 n(m) = \int {\rm d}{\bmath h}\,p({\bmath h})\,{\cal C}_m({\bmath h}),
\ee
where ${\cal C}_m({\bmath h})$ specifies the set of constraints on ${\bmath h}$ which must be satisfied to form a halo of mass $m$.  Then 
\be
 n(m|\Delta_e) = \int {\rm d}{\bmath h}\,p({\bmath h}|\Delta_e)\,
                      {\cal C}_m({\bmath h})
\label{nme}
\ee
and 
\be
 \langle\Delta_0|\Delta_e,m\rangle = 
 \frac{\int {\rm d}{\bmath h}\,p({\bmath h},\Delta_e)\,{\cal C}_m({\bmath h})\,\langle\Delta_0|\Delta_e,{\bmath h}\rangle}{\int {\rm d}{\bmath h}\,p({\bmath h},\Delta_e)\,{\cal C}_m({\bmath h})}.
 \label{general}
 \ee

\subsection{Reconstructing the dependence on mass and environment}
\label{sec:brec}
Fixing $m$ and marginalizing over all $\Delta_e$ yields 
\begin{align}
 & \frac{\int {\rm d}\Delta_e \int {\rm d}{\bmath h}\,p({\bmath h},\Delta_e)\, {\cal C}_m({\bmath h})\,\langle\Delta_0|\Delta_e,{\bmath h}\rangle}{\int {\rm d}\Delta_e \int {\rm d}{\bmath h}\,p({\bmath h},\Delta_e)\,{\cal C}_m({\bmath h})} \nonumber\\
 &= \frac{\int {\rm d}{\bmath h}\,p({\bmath h})\,{\cal C}_m({\bmath h})\,\langle\Delta_0|{\bmath h}\rangle}{\int {\rm d}{\bmath h}\,p({\bmath h})\,{\cal C}_m({\bmath h})\int {\rm d}\Delta_e\,p(\Delta_e|{\bmath h})} \nonumber\\
 & \quad + \ \frac{\int {\rm d}{\bmath h}\,p({\bmath h})\,{\cal C}_m({\bmath h})\,\int {\rm d}\Delta_e \, p(\Delta_e|{\bmath h})\,\langle\Delta_0|\Delta_{e|{\bmath h}}\rangle}{n(m)} \nonumber\\
 &= \frac{\int {\rm d}{\bmath h}\,p({\bmath h})\,{\cal C}_m({\bmath h})\,\langle\Delta_0|{\bmath h}\rangle}{n(m)}.
 \label{brecm}
\end{align}
The ratio of the final expression to $\langle\Delta_0\Delta_h\rangle$ is what is usually meant by $b_h(m)$.  

On the other hand, marginalizing over all halo masses at fixed environment yields 
\begin{align}
 \label{brecd}
  & \frac{\int {\rm d}m \int {\rm d}{\bmath h}\,p({\bmath h}|\Delta_e)\, {\cal C}_m({\bmath h})\,\langle\Delta_0|\Delta_e,{\bmath h}\rangle}{\int {\rm d}m \int {\rm d}{\bmath h}\,p({\bmath h}|\Delta_e)\,{\cal C}_m({\bmath h})} \\
 &= \langle\Delta_0|\Delta_e\rangle + \frac{\int {\rm d}m \int {\rm d}{\bmath h}\,p({\bmath h}|\Delta_e)\, {\cal C}_m({\bmath h})\,\langle\Delta_0|\Delta_{{\bmath h}|e}\rangle}{\int {\rm d}m\,n(m|\Delta_e)}  \nonumber.
\end{align}
If ${\bmath h}$ involves $\Delta_h$ (where, typically $R_h^3\propto m$) and its derivatives, then, as we have already discussed, we expect the expression above to be dominated by the first term on the right hand side, especially when $R_h\ll R_e\ll R_0$.  

\citet{Pujol2017} show that if one attempts to reconstruct how bias depends on $\Delta_e$ using 
\begin{equation*}
 \frac{\int {\rm d}m \int {\rm d}{\bmath h}\,p({\bmath h}|\Delta_e)\, {\cal C}_m({\bmath h})\,\langle\Delta_0|{\bmath h}\rangle}{\int {\rm d}m\,n(m|\Delta_e)} 
\end{equation*}
(their equation~5) then one gets the wrong answer:  almost no predicted dependence of the bias on $\Delta_e$ when the measurements show a strong trend.  Comparison with our equation~(\ref{brecd}) shows why; by assuming that halo bias depends only on halo mass, their expression misses the contribution which leads to the first term on the right hand side of our expression -- the term which dominates the answer when $R_m\ll R_e\ll R_0$.  

On the other hand, \citet{Pujol2017} found that 
\begin{equation*}
 \frac{\int {\rm d}\Delta_e \int {\rm d}{\bmath h}\,p({\bmath h},\Delta_e)\, {\cal C}_m({\bmath h})\,\langle\Delta_0|\Delta_e\rangle}{\int {\rm d}\Delta_e\,n(\Delta_e|m)} 
\end{equation*}
(their equation~6) was able to reconstruct the mass dependence of bias rather well.  Our analysis shows why this works, even though it too is, formally, incorrect.  (The correct expression is our equation~\ref{brecm}.)  Namely, the expression above can be written as   
\begin{align}
 & \frac{\int {\rm d}{\bmath h}\,p({\bmath h})\, {\cal C}_m({\bmath h}) \int {\rm d}\Delta_e \,p(\Delta_e|{\bmath h})\,\langle\Delta_0|\Delta_e\rangle}{n(m)} \nonumber\\
 &= \frac{\int {\rm d}{\bmath h}\,p({\bmath h})\, {\cal C}_m({\bmath h}) \,\langle\Delta_0\Delta_e\rangle \langle\Delta_e|{\bmath h}\rangle/S_e}{n(m)} ,
\end{align}
and our discussion of equation~(\ref{orth}) showed that we expect 
\be
 \langle\Delta_0\Delta_e\rangle \langle\Delta_e|{\bmath h}\rangle/S_e \approx 
 \langle\Delta_0|{\bmath h}\rangle .
\ee
If this approximation were an equality, then their expression would reduce to the correct one, our equation~(\ref{brecm}).  That it is only an approximation is why \citet{Pujol2017} only found good, but not perfect agreement with the actual mass dependence of bias, $b_h(m)$.  

\section{Evolution}\label{Eul}
The analysis of the previous section was for statistics in the initial conditions, sometimes called Lagrangian space.  Since the analysis in \citet{Pujol2017} was for halos and environments defined in the evolved Eulerian space, our assertions in Section~\ref{Lag} are not completely justified until we have shown that they survive nonlinear evolution.  

\subsection{Excursion set approach:  Analytic}
We use the excursion set approach of \citet{Sheth1998} to model statistics in the evolved Eulerian space.  This approach makes use of the spherical evolution mapping between $\delta_{\rm V}$, the Eulerian density on scale $V$, and $\Delta_{\rm M}$, the Lagrangian density on scale $M$:
\be
 1 + \delta_{\rm V} \equiv M/\bar\rho V
 = (1 - \Delta_{\rm M}/\delta_c)^{-\delta_c} ,
 \label{scapprox}
\ee
where $V$ is the Eulerian volume, $M$ is the mass in it, and $\delta_c\approx 1.686$ (although $\delta_c = 21/13$ reproduces the monopole of second order perturbation theory).  In what follows, we will also make use of the fact that
\begin{equation}
  \frac{\Delta_{\rm M}}{\delta_c} = 1 - \left(\frac{\bar\rho V}{M}\right)^{1/\delta_c}
               = 1 - (1+\delta_{\rm V})^{-1/\delta_c},
  \label{scbarrier}
\end{equation}
which follows from rearranging equation~(\ref{scapprox}).

The gist of the argument is that Eulerian statistics on scale $V$ are related to Lagrangian statistics on scale $M$.  While this idea can be traced back to \citet{Bernardeau1994}, the analysis in \citet{Sheth1998} allows one to work down to substantially smaller $V$.  \citet{Lam2008} show that it provides a rather good model of what we call the probability distribution of the Eulerian environment here.  Our goal is to show that this approach also provides a simple description of the joint distribution of halos and their environment -- i.e. of Eulerian bias -- a point which was made in \citet{Sheth1998}, but has not been followed-up since.  This turns out to be straightforward, particularly because of recent advances in our understanding of the excursion set approach \citep{Musso2012}.

In what follows, $\delta$ and $\Delta$ always denote Eulerian and Lagrangian overdensities, and their subscripts always denote the corresponding Eulerian or Lagrangian smoothing scale.  E.g., if $\delta_0$ is the Eulerian density on scale $V_0$, then $\Delta_0$ is the Lagrangian density on scale $M_0/\bar\rho = V_0\,(1 + \delta_0)$, and $\delta_0$ and $\Delta_0$ are related by equation~(\ref{scbarrier}).

Our goal is to estimate the mean Eulerian density on scale $V_0$ given that the Eulerian cell is centred on a region with Eulerian density $\delta_e$ on scale $V_e$ which itself is centred on a halo of mass $m$.  The Lagrangian version of this quantity is equation~(\ref{general}).  It becomes 
\be
 \langle\delta_0|\delta_e,m\rangle =
 \frac{\int {\rm d}{\bmath h}\,p({\bmath h},\Delta_e)\,{\cal C}_m({\bmath h})\,\langle\delta_0|\Delta_e,{\bmath h}\rangle}{\int {\rm d}{\bmath h}\,p({\bmath h},\Delta_e)\,{\cal C}_m({\bmath h})},
 \ee
 where we have used the fact that the $m$ and $\delta_e$ constraints correspond to simple constraints in Lagrangian space. The main problem is to estimate $\langle\delta_0|\Delta_e,\bmath{h}\rangle$.  

On large Eulerian scales $V_0$ we expect $\delta_0\ll 1$, and hence $\Delta_0\approx \delta_0$ almost surely.  In this limit, we expect to be able to use the Gaussian expression (equation~\ref{orth}):
\begin{equation}
  \langle\delta_0|\delta_e,m\rangle 
  \approx \langle\Delta_0|\Delta_e\rangle + \langle\Delta_0|\Delta_{h|e}\rangle
 \label{2terms}
\end{equation}
where $\langle\Delta_0|\Delta_e\rangle = \Delta_e\,\langle\Delta_0\Delta_e\rangle/\langle\Delta_e^2\rangle$ dominates.  This would make
\begin{equation}
  b_{\rm e}^{\rm E} = \frac{\Delta_e}{\langle\Delta_e^2\rangle}
  = \frac{\delta_c[1 - (1+\delta_e)^{-1/\delta_c}]}{S[\bar\rho V_e (1+\delta_e)]},
  \label{kaiserE}
\end{equation}
where $S$ is the the Lagrangian variance on the mass scale $\bar\rho V_e (1+\delta_e)$. Equation~(\ref{kaiserE}) is the expression in \citet{Abbas2007}.  Comparison with equation~(\ref{bKaiser}) shows explicitly that, in this limit, the Eulerian bias is like the Lagrangian one provided that one correctly rescales the density {\em and} volume.  Figure~3 of \citet{Pujol2017} shows that this simple expression works remarkably well over a wide range of scales.

Before moving on, we note that this expression has been rediscovered by \citet{Uhlemann2017} who appear to be unaware of earlier work.  Moreover, as we have spelled out in more detail here, the excursion set approach of \citet{Sheth1998} shows why, even though equation~(\ref{kaiserE}) is quite accurate, it is just an approximation.  That is to say, it shows clearly how to go beyond rescaled Kaiser-bias. For example, the top panel of Figure~6 in \citet{Pujol2017} shows the comoving number density of halos in cells of specified overdensity $\delta_e$.  They do not remark on it, but this quantity has long been known to be well-approximated by equation~(\ref{nme}), with $\Delta_e$ given by $\delta_c$ times the rhs of equation~(\ref{scbarrier}) when $\delta_e=\delta_V$ in the rhs of equation~(\ref{scbarrier}) \citep{Mo1996, Sheth2002}.  This is a limit which our approach is designed to reproduce \cite{Sheth1998}. 

\begin{figure}
 \centering
 \includegraphics[width=1.\hsize]{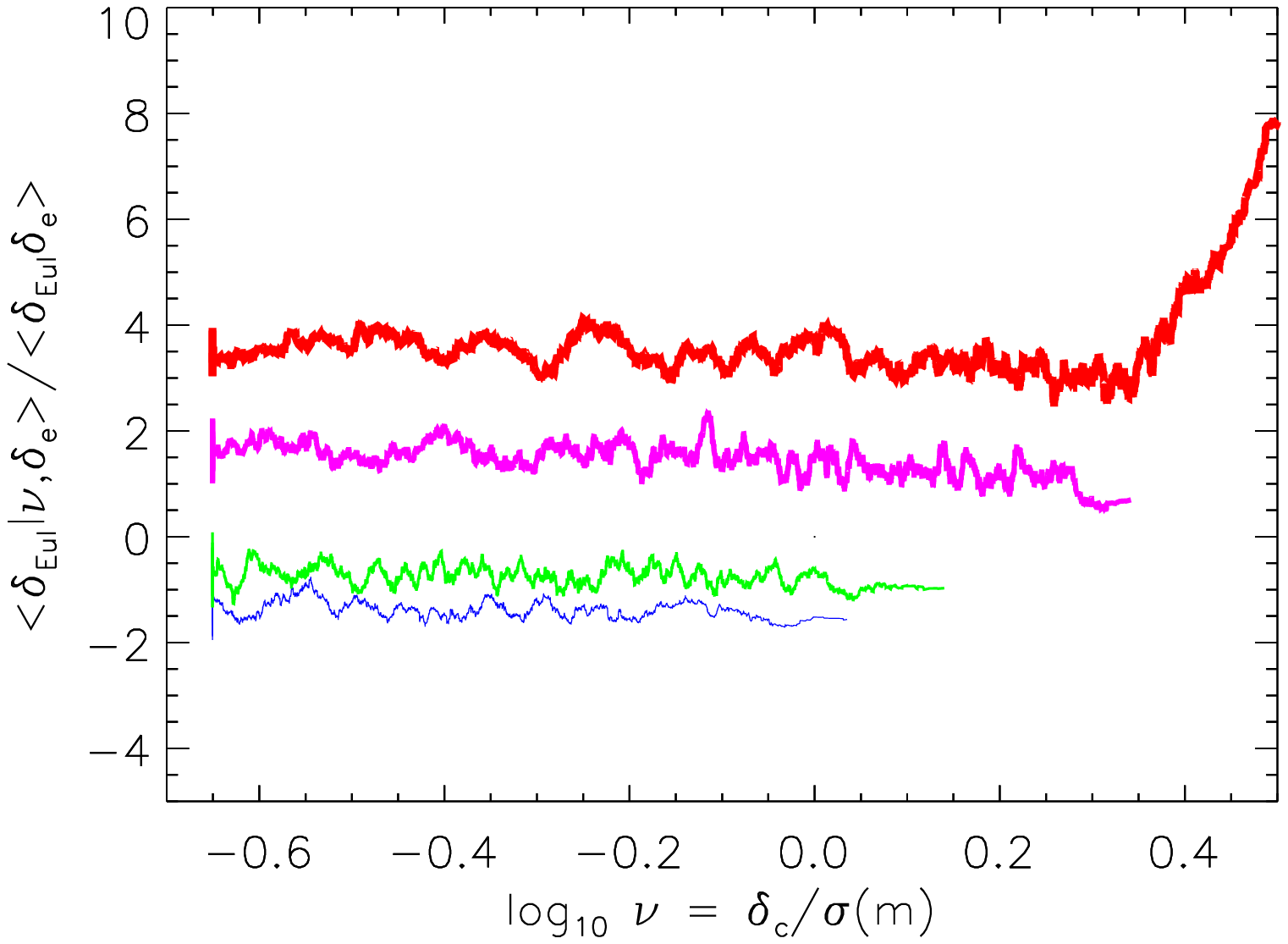}
 \caption{The Eulerian bias of halos surrounded by large overdensities is larger; however, at fixed overdensity, bias is the same for all except the most massive halos. The thickness of line is proportional to the overdensity value, with the thickest line corresponding to the densest field. Red and magenta lines are for halos in the densest 10\% and the next densest 20\% of the cells in the evolved Eulerian field; blue and green lines show the bias of halos in the least dense 10\% and the next emptiest 20\%.   The apparent upper limit in $\nu$, which increases with density, is because massive halos are not present in the least dense cells.}  
 \label{bMfixE}
\end{figure}

\subsection{Excursion set approach: Monte-Carlo}
We have checked the analysis above explicitly in Monte-Carlo realizations of this process.  Namely, we generated $10^5$ random walks, each having a correlation structure appropriate for tophat smoothing of Gaussian field having $P(k)\propto k^{-2}$.  We used the algorithm described in \citet{Musso2014} to do this.  For each walk, we stored the mass scale on which it first crossed a `constant barrier' of height $\delta_c$, and the mass scale on which it crossed the `moving barrier' of equation~(\ref{scbarrier}), for a range of choices of Eulerian $V$.  First crossing of $\delta_c$ is a simple proxy for a halo; by storing first crossings for a range of $V$, we can map out the Eulerian profile around each `halo' (see \citealt{Sheth1998}; indeed, viewed this way, a halo is just the special case in which $V=0$).  In addition, we stored the height of the walk on a number of mass scales, which we use to reconstruct Lagrangian profiles of halos or of Eulerian cells.  

\begin{figure}
 \centering
 \includegraphics[width=1.\hsize]{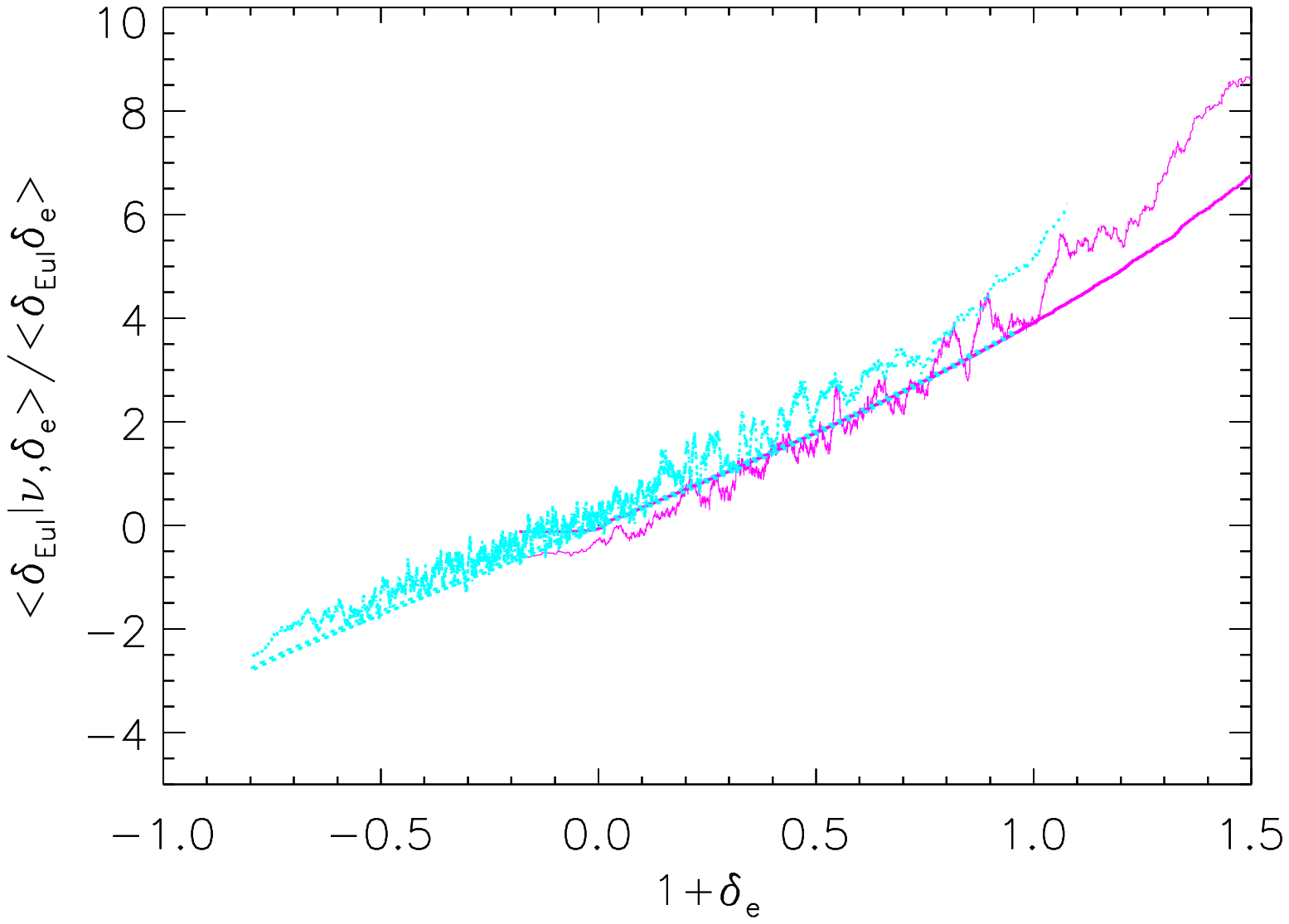}
 \caption{Denser Eulerian cells are more biased, but this bias is independent of the mass of the halo at the cell centre.  Cyan dotted and magenta solid lines show results for cells centred on the 10\% lowest and highest mass halos.  Thick smooth curve shows $b_e^{\rm E}$ of equation~(\ref{kaiserE}). }
 \label{bEfixM}
\end{figure}

Red, magenta, green and blue curves in Figure~\ref{bMfixE} show the Eulerian bias of halos which are centred on patches having Eulerian densities $\langle 1+\delta_{\rm V}\rangle = (7.5, 2.3, 0.5, 0.3)$.  The scale $V$ is such that, when smoothed on scale containing mass $M=\bar\rho V$, the rms linear theory overdensity had variance $\langle\Delta_{\rm V}^2\rangle =0.5^2$.  (Therefore, a halo of mass $\bar\rho V$ would have $\nu = \delta_c/\sigma = 42/13$.  This is why the curves for underdense regions do not extend to larger $\nu$.)  Clearly, the Eulerian bias is larger for the halos centred on denser cells; however, except for the densest cells, the bias is the same for all halo masses.  I.e., the bias is determined by the environment, and not by halo mass.

Figure~\ref{bEfixM} shows another way of presenting this trend:  cyan and magenta curves show how the bias depends on environment for the least and most massive halos.  Clearly, the bias is the same strong function of environment whatever the mass of the halo at the centre.  (For the moment, we are ignoring the slight tendency for the cells centred on the most massive halos to have slightly smaller bias factors.)  The smooth curve shows equation~(\ref{kaiserE}); it provides a good description of the measurements.  This shows explicitly that the first term in equation~(\ref{2terms}) really does capture most of the environmental effect.  The second term in equation~(\ref{2terms}) must account for the small trend with mass which remains, but note that this is much smaller than the overall trend with environment.  

\begin{figure}
 \centering
 \includegraphics[width=0.9\hsize]{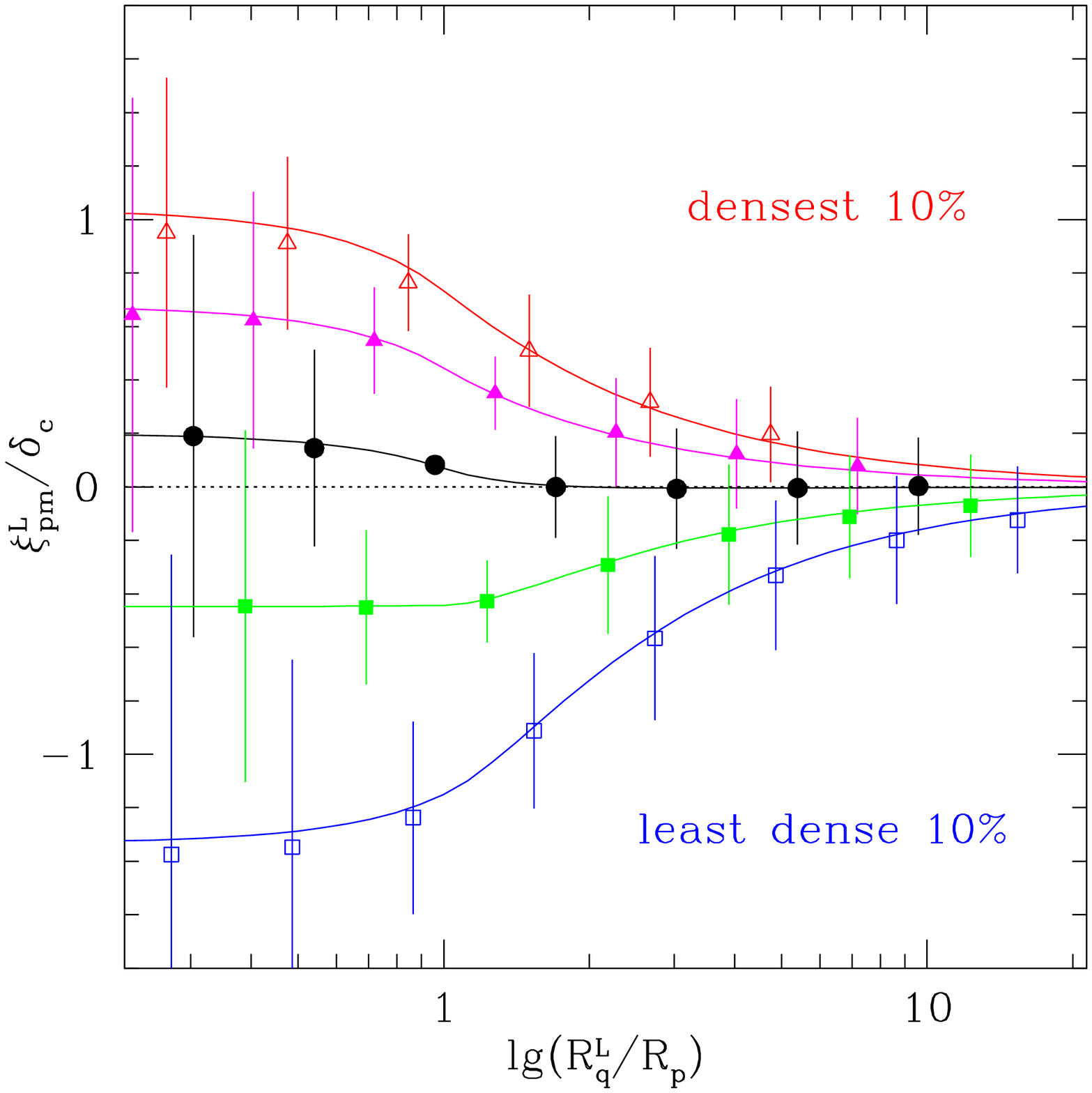}
 \caption{Initial Lagrangian density profiles around patches which become Eulerian cells with the specified density:  red and magenta are for the densest 10\% and the next densest 20\% of the cells in the evolved Eulerian field; blue and green show the least dense 10\% and the next emptiest 20\%.  Symbols with error bars show the mean and the rms around the mean -- errors on the mean are smaller than the symbols.  Curves show the Lagrangian-space cross correlation; there are no free parameters in this comparison.  }
 \label{xipmL}
\end{figure}

The trends in Figures~\ref{bMfixE} and~\ref{bEfixM} are remarkably similar to those shown in Figure~4 of \citet{Pujol2017}.  Even the slight tendency for environments centred on massive halos to be slightly less biased (Figure~\ref{bEfixM}) is similar.  This suggests that our Monte Carlos have captured the essence of the effect.

To show that we really do understand the origin of this effect, on small scales as well, Figure~\ref{xipmL} shows the initial Lagrangian density profiles around the patches which evolve into the densest and least dense cells.  Symbols with error bars show the mean and the rms around the mean -- errors on the mean are smaller than the symbols.  Curves show the Lagrangian-space cross correlation -- essentially equation~(\ref{orth}); there are no free parameters in this comparison. The agreement justifies the assertions we made in Section~\ref{sec:brec}.  Namely, accounting for the joint distribution of mass and environment is straightforward.  Doing so shows that, just as in Lagrangian space, the Eulerian bias is also determined primarily by the larger scale environment, and much less so by halo mass.  Therefore, analyses which ignore the environmental effect will lead to incorrect conclusions about the nature of halo bias.

\section{Discussion and conclusions}
We discussed how the large scale bias of halos depends on both halo mass and environment, in Lagrangian (Section~\ref{Lag}) and Eulerian (Section~\ref{Eul}) space.  We showed that, at fixed environment, the dependence of large scale bias on halo mass should be weak (Figure~\ref{bMfixE}).  Indeed, if one defines halos and environment using a filter which is sharp in $k$-space \citep{Bond1991}, and one conditions on large scale environment, then large-scale bias is predicted to be {\em completely} independent of halo mass (see discussion in Section~\ref{sharpk}).  Our calculation quantifies the small residual effect which comes from the fact that correlations between scales are more complicated than for sharp-$k$ smoothing.

This has an interesting implication.  Following \citet{Sheth2004}, there have been many studies of the dependence of bias on other parameters, if halo mass is held fixed.  These are usually called `Assembly Bias' studies, even though the additional parameters may not be explicitly related to halo assembly.  The underlying origin of all these signals is nontrivial correlations between scales.  Our analysis shows that if bias at fixed environment does show some dependence (presumably weak!) on halo mass, then one has detected the effect of nontrivial correlations between scales.  In this sense, one has detected `Assembly Bias' coming from the other way round from what is currently fashionable.  To see this explicitly, suppose we order scales as
  $$R_{\rm bias} \ge R_{\rm env} \ge R_{\rm halo} \ge R_{1/2}$$
where $R_{1/2}$ (for `half-mass') is a crude proxy for halo assembly.  The usual studies fix $R_{\rm halo}$ and look for additional correlation between the more widely separated scales $R_{\rm bias}$ and $R_{1/2}$.  But the ranking of scales shows that one could have fixed $R_{\rm env}$ and looked for additional correlation between $R_{\rm bias}$ and $R_{\rm halo}$ (or $R_{1/2}$).  This is the sense in which looking for mass dependence at fixed environment is the same as assembly bias.

The analysis of the previous section is particularly relevant to the question of whether or not galaxy properties depend on quantities other than halo mass. The main text shows why, when the environment is constrained, then halo bias is a function of both halo mass and environment.  However, this does {\em not} mean that the Halo Occupation Distribution of how galaxies populate halos must also depend on both (it may, but it need not).  Indeed, \citet{Abbas2007} showed that mock galaxy catalogs, in which mass is the only variable which determines how galaxies populate halos, automatically exhibit a number of environmental trends that are seen in the data.  That is to say, they showed that the data they examined do {\em not} require any additional galaxy-environment effect:  the halo-environment correlation which comes for free, and which we have spelled out in some detail in this paper, is sufficient to explain the galaxy-environment correlations.

While this may be true for observables such as luminosity, which are expected to be monotonically related to halo mass, the same may not be true for colors, for which the correlation with halo mass is not as simple.  A simple model for galaxy colors, in which galaxy-environment correlations are inherited from the halo-environment correlations, is able to provide a reasonable description of the bright SDSS galaxies considered by Abbas \& Sheth \citep{Skibba2009}.  However, it is too simplistic to account for all observed correlations \cite[see][for the current state of the art]{Pahwa2017}, and studies at the faint end have yet to be done.  \citet{Pujol2017} show that, in the semi-analytic galaxy formation model they considered, galaxy color appears to correlate more with density than halo mass, and that density appears to be more important than halo mass for faint red central galaxies.  The importance of environment over halo mass appears at lower luminosities than Abbas \& Sheth considered in the SDSS.  At these lower luminosities, the scatter between halo mass and luminosity becomes larger, so it will be interesting to see if the color-dependent trends in \citet{Pujol2017}'s Figure~8 are reproduced in data.  With such studies in mind, we express their results in our formalism in Appendix~\ref{pujol}.

An interesting extension of our work would be to study what happens if the environment of a halo is defined using a measure which does not correlate with the density.  E.g., \citet{Paranjape2017} use a measure which is built from the tidal shear.  They show that, at fixed mass, halo bias correlates strongly with the morphology of the environment (e.g., `filamentary' versus `isotropic'), {\em and} that bias is also a strong function of mass when the environment is fixed.  This `assembly bias' effect appears to be richer than the one with density which we studied here.  Again, however, galaxies will inherit the environmental correlations of their host halos, so care must be taken to isolate correlations with environment which are over and above those which come `for free' from the host halo-environment correlation.

Finally, the careful reader will have noticed that our least dense cells have $1+\delta_e\sim 0.2$ (e.g. Figure~\ref{bMfixE}); such cells would be classified as `voids' \citep{ShethWeygaert2004}.  These `voids' have Eulerian bias factors which are less than zero (Figure~\ref{bMfixE}), and the associated Lagrangian profiles of these cells are indeed rather underdense, especially on small scales (Figure~\ref{xipmL}).  Clearly, then, the excursion set approach allows us to model the evolution of void profiles; this is done in Massara \& Sheth (2017, in preparation). 
In addition, study of the redshift-space clustering in the $b=0$ subsample (Figures~\ref{xi3dx} and~\ref{xi3db} suggest this is true of the 30\% underdense sample) may allow simple constraints on the growth rate $f = {\rm d}\ln D/{\rm d}\ln a$, from a comparison of the (projected) real and redshift space clustering signals.  Furthermore, subsamples selected using our methodology have a rather wide range of bias factors, making them well-suited for multi-tracer constraints on redshift space distortions and primordial non-Gaussianity \citep{McDonald2009}, and for measuring the gravitational redshift effect from large scale structures \citep{Zhu2017}. 

\section*{Acknowledgements}
We thank the ICTP for its hospitality during the summers of 2015, 2016 and 2017, Nordita for its hospitality during the program `Advances in Theoretical Cosmology in Light of Data' in July 2017, and A. Pujol, E. Gazta\~naga and A. Paranjape for helpful discussions. 

\bibliographystyle{mnras}
\bibliography{biasEnv}

\appendix
\section{Relation to the work of Abbas \& Sheth (2007)}

It is natural and common to define the environment of a galaxy by counting the number of other galaxies within a specified distance from it.  Suppose that the scale which defines the environment is substantially larger than the galaxy itself, and one selects a subset of galaxies based on this environment.  If one measures the galaxy-galaxy correlation function for each such subset, then one will find that galaxy clustering is not a monotonic function of environment \citep{Abbas2007}.  Whereas the galaxies with the most neighbours are the most strongly clustered, those with the fewest neighbours are not the least strongly clustered:  the least clustered galaxies are associated with only moderately underdense environments.  

\citet{Abbas2007} showed that this effect was present in a mock catalog in which the number of galaxies in a halo depends on halo mass and not its environment.  The agreement between the enviromental trends in the data and in their mock catalog means that we can make other measurements, some of which are not possible in the data, so as to illustrate a few other interesting points.

\begin{figure}
 \centering
 \includegraphics[width=0.9\hsize]{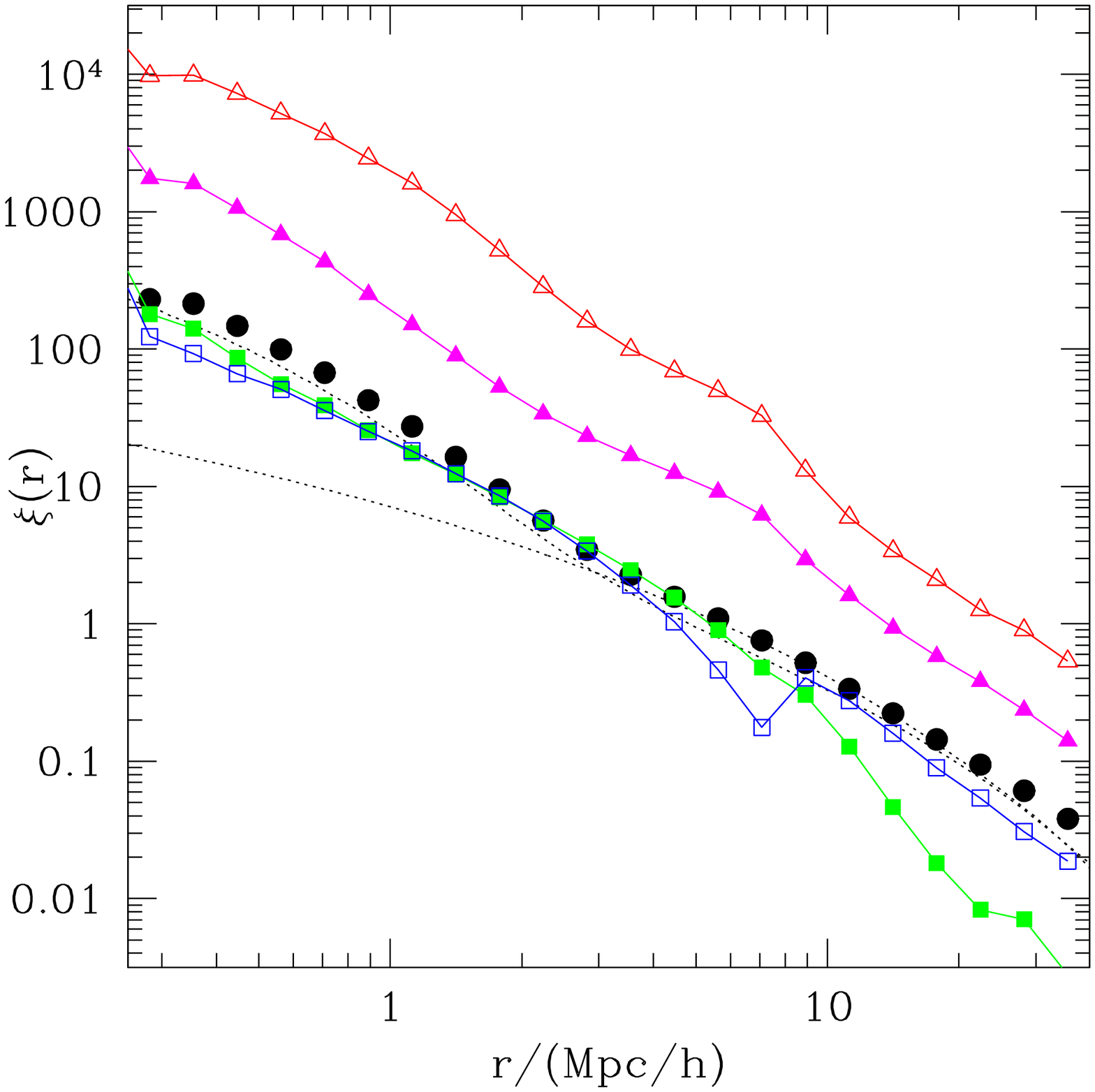}
 \caption{Environmental dependence of the galaxy-galaxy auto-correlation function.  Results for four bins in environment -- defined to be the number of galaxies within $8h^{-1}$Mpc -- are shown.  Open triangles, filled triangles, filled squares and open squares show results for the densest to the least dense environments.  This measure of clustering is not a monotonic function of environment.  To guide the eye, filled circles show the auto-correlation function of the full sample, and the two dotted curves show the dark matter correlation function in linear and nonlinear theory.  }
 \label{xi3dd}
\end{figure}

To reduce the effect of redshift space distortions, the measurements in the data were restricted to a projected measurement.  Figure~\ref{xi3dd} shows the corresponding real-space measurement in the mock catalog.  Galaxies were ranked by the number of objects within $8h^{-1}$Mpc; empty triangles show the clustering signal for the objects in the top ten percentile, and filled triangles show it for the objects between the top ten and thirty percentiles.  To help set the scale, the dotted lines show $\xi_{mm}$ in linear theory and nonlinear theory (larger on small scales).  Open and filled squares show the objects in the bottom ten, and between the bottom ten and thirty percentiles.  Notice the effect mentioned above:  clustering is not a monotonic function of environment.  In particular, the objects in moderately underdense patches are very weakly clustered (filled squares).

\begin{figure}
 \centering
 \includegraphics[width=0.9\hsize]{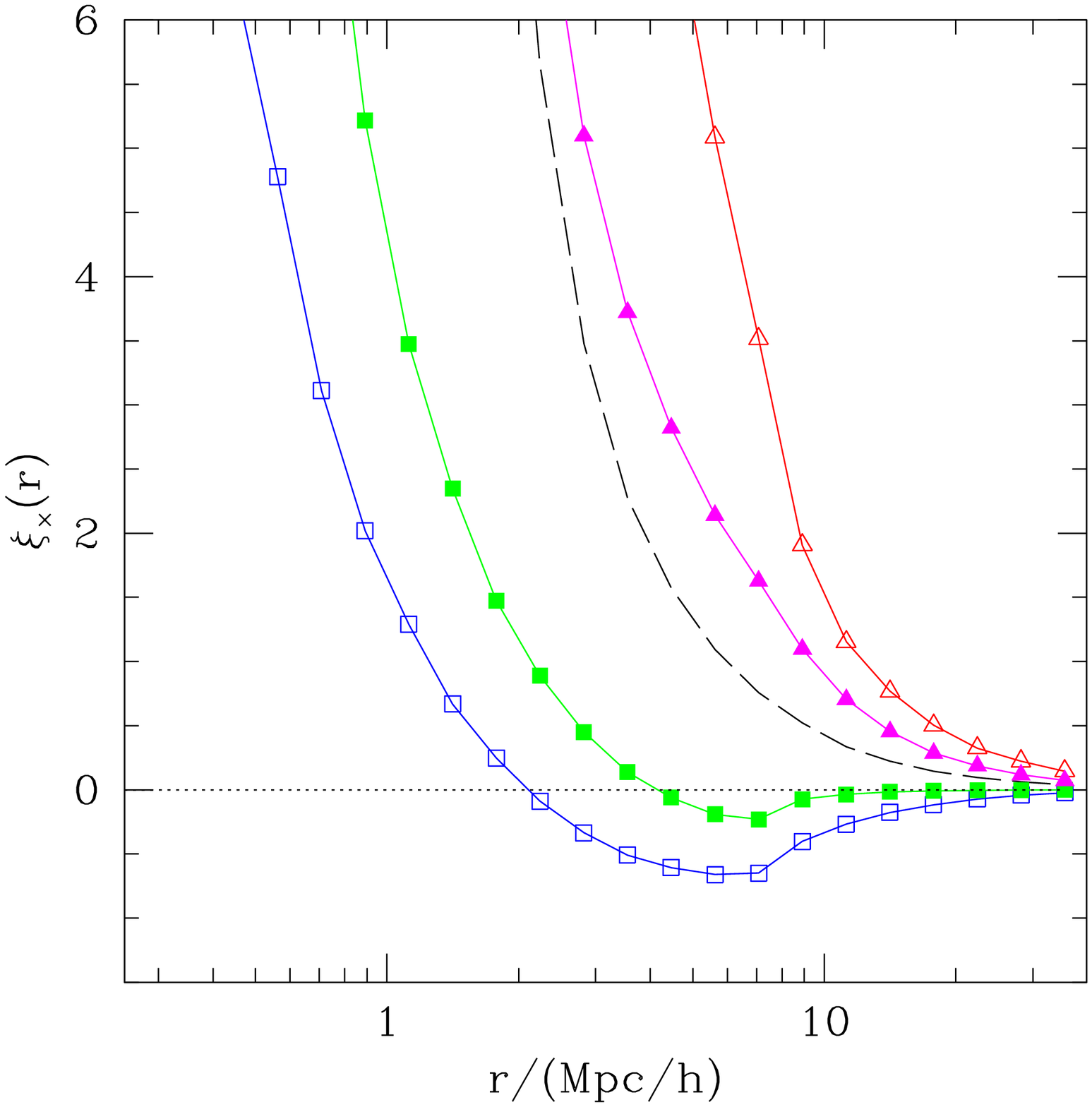}
 \caption{Environmental dependence of the galaxy-total cross-correlation function.  The symbols show the same environmental bins as before, cross-correlated with the full sample (dashed curve shows the auto-correlation function of the full sample).  Note that the y-axis is linear rather than log, since the signal for underdense regions crosses zero.  This signal is clearly monotonic with environment.}
 \label{xi3dx}
\end{figure}

This non-monotonicity is a consequence of the fact that the measurement is an auto-correlation function.  On large scales, $\xi_{gg}\propto b_{g}^2\,\xi_{mm}$, so the measurement cannot distinguish between positive and negative $b_g$.  While this was implicit in their discussion, \citet{Abbas2007} did not show a plot illustrating that $b_g$ itself {\em is} monotonic with environment (and, in particular, is negative for underdense regions).  To rectify this omission, Figure~\ref{xi3dx} shows the cross-correlation between each subsample and the total.  This signal is clearly monotonic with environment.

\begin{figure}
 \centering
 \includegraphics[width=0.9\hsize]{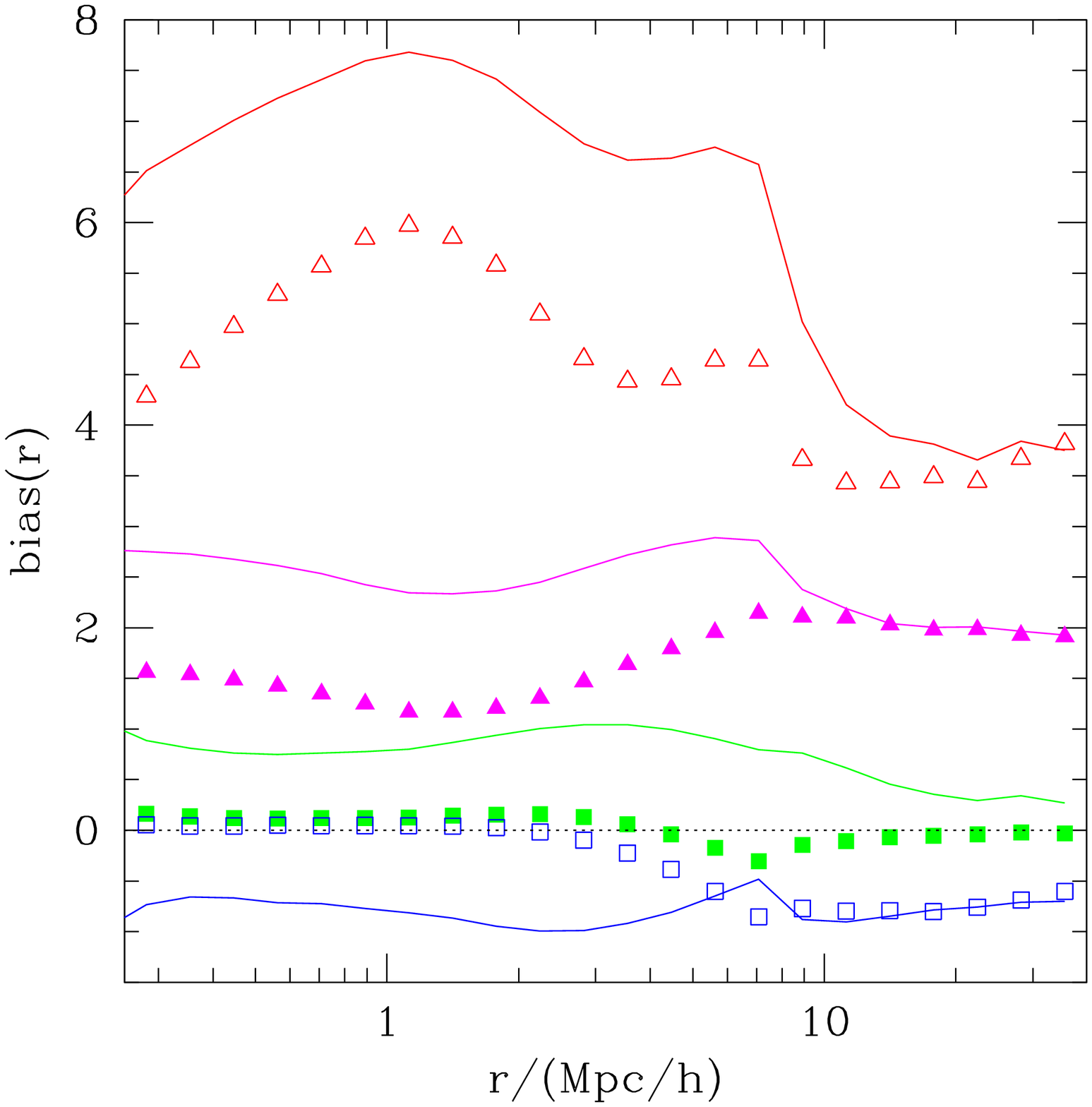}
 \caption{Environmental dependence of galaxy bias.  Symbols show the ratio of the cross-correlation to the auto correlation of the full sample (symbols divided by dashed curve in previous figure); curves show the square-root of the ratio of the ratio of the symbols divided by filled circles in Figure~\ref{xi3dd} (and multiplied by $-1$ for the least dense region).}
 \label{xi3db}
\end{figure}

If we use $b_t$ to denote the bias factor of the full sample and $b_g$ that of a subsample, then, on large scales, this cross-correlation signal should be proportional to $\xi_{gt}\propto b_gb_t\xi_{mm}$.  Since $b_t$ is the same for all the measurements, the amplitude of the signal on large scales is proportional to $b_g$.  To highlight this, the symbols in Figure~\ref{xi3db} show the ratio $\xi_{gt}/\xi_{tt} = (b_g/b_t)$ for the four subsamples (while this does not matter for our argument, it may help to notice from Figure~\ref{xi3dd} that $b_t\sim 1$).  The symbols show clearly that $b_g$ is monotonic with environment:  it is negative for the least underdense regions, close to zero for moderately underdense regions, and positive for overdense regions.  

To show that the non-monotonic signal in the auto-correlation function is a consequence of it not being able to distinguish between positive and negative bias factors, the lines show $\sqrt{\xi_{gg}/\xi_{tt}}$, except for the signal from the least dense regions, which we multiply by $-1$.  Except for the sample which had $b_g\sim 0$, the agreement with the symbols is excellent on large scales indicating that, indeed, the non-monotonicity in $\xi_{gg}$ is because it scales as $b_g^2$.  

The main text develops a model for the precise trend with environment.  It is interesting that the bins in environment here lead to rather similar large scale bias as shown in Figure~\ref{bMfixE} of the main text.

\section{Implications for galaxy bias}\label{pujol}

Figure~8 in \cite{Pujol2017} shows that a halo mass based approach (mHOD) does not recover the color dependence of galaxy bias, while a density based model (dHOD) does.  Below, we express their results in our notation.

Consider the case of bias as a function of color.  For Gaussian distributions, the mean overdensity $\Delta_0$ at fixed color is 
\be
 \langle\Delta_0 | g_c \rangle =
 \langle\Delta_0 g_c\rangle\frac{g_c}{\langle g_c^2\rangle}\,\,,
\label{eq:bias_color}
\ee
where $g_c$ stands for the constraint which specifies galaxy color.  The mHOD model approximates this as 
\be
 \int {\rm d}m\, \langle\Delta_0 | m\rangle\, p(m | g_c)
 = \frac{\langle \Delta_0 m\rangle \langle mg_c \rangle}{\langle m^2\rangle}\frac{g_c}{\langle g_c^2 \rangle}
 \label{eq:biasc_rec_m}
\ee
(their equation 5).  This will be a good approximation if $\langle \Delta_0 m\rangle \langle mg_c \rangle/\langle m^2\rangle \approx \langle\Delta_0 g_c\rangle$:  i.e., if the $\Delta_0$-color correlation is entirely due to the correlations of each with halo mass $m$.  The grey curve in the upper panel of their Figure~8 shows that this is a poor approximation to the actual relation; mass alone cannot account for the $\Delta_0$-color correlation.

The dHOD approximation uses the environment $\Delta_e$ instead of halo mass:  
\be
 \int {\rm d}\Delta_e\, \langle\Delta_0 | \Delta_e\rangle \, p(\Delta_e | g_c)
 = \frac{\langle \Delta_0 \Delta_e\rangle \langle \Delta_e g_c \rangle}{\langle \Delta_e^2\rangle}\frac{g_c}{\langle g_c^2 \rangle}
 \label{eq:biasc_rec_d}
\ee
(their equation 6).  The red curve in the upper panel of their Figure~8 shows that this works well, implying that $\langle \Delta_0 \Delta_e\rangle \langle \Delta_e g_c \rangle/\langle \Delta_e^2\rangle \approx \langle\Delta_0 g_c\rangle$; the $\Delta_0$-color correlation is almost entirely due to the correlations of each with $\Delta_e$.

The case of bias for given luminosity is similar to the one for given color by simply replacing $g_c$ with $g_{\rm L}$, where $g_{\rm L}$ represents galaxy luminosity. (Strictly speaking, to model the bottom panel in their Figure~8 $g_{\rm L}$ must stand for the luminosity of a red central galaxy.)  At large luminosities, there is a tight $m-g_{\rm L}$ relation, so the mHOD approach (grey curve in their lower panel) provides a good approximation to the true relation.  However, at the faint end, the $m-g_{\rm L}$ relation is much looser, so the mHOD reconstruction fails.  As there are many more faint red galaxies than bright, the statistics in their top panel are dominated by the faint objects, for which the mHOD approach fails.

\label{lastpage}

\end{document}